\documentclass[a4paper,10pt,twocolumn]{article}
\usepackage{graphicx,amsmath,jpc,amssymb}
\begin{document}

\title{Heat conductance is strongly anisotropic for pristine silicon nanowires}

\author{Troels Markussen$^1$, Antti-Pekka Jauho$^{1,2}$, Mads Brandbyge$^1$\\$^1$ Department of Micro- and Nanotechnology\\ Technical University of Denmark, DTU Nanotech\\ Building 345 East, DK-2800 Kgs. Lyngby, Denmark\\
$^2$ Laboratory of Physics, Helsinki University of Technology\\ P.O. Box 1100, FIN-02015 HUT, Finland
}

\date{\today}

\maketitle

\begin{abstract}
We compute atomistically the heat conductance for ultra-thin
pristine silicon nanowires (SiNWs) with diameters ranging from 1 to
5 nm. The room temperature thermal conductance is found to be
highly anisotropic: wires oriented along the  $\langle110\rangle$ direction
have 50--75\% larger conductance than wires oriented along the
$\langle100\rangle$ and $\langle111\rangle$ directions. We show that
the anisotropies can be qualitatively understood and reproduced from
the bulk phonon band structure. \textit{Ab initio} density
functional theory (DFT) is used to study the thinnest wires, but
becomes computationally prohibitive for larger diameters, where we
instead use the Tersoff empirical potential model (TEP). For the
smallest wires, the thermal conductances obtained from DFT- and TEP
calculations agree within 10\%. The presented results could be
relevant for future phonon-engineering of nanowire devices.

\end{abstract}

\vspace{1cm}

Thermal transport properties of silicon nanowires (SiNWs) are
crucial for many of their future applications. A high thermal
conductance of SiNWs is desirable in nano-electronic components,
while, on the other hand, a low thermal conductance is beneficial
for thermoelectric applications. Indeed, recent experiments have
demonstrated that SiNWs can have a significantly higher
thermoelectric performance than bulk silicon
\cite{HochbaumNature2008,BoukaiNature2008}, due to a reduced thermal
conductance of the nanowires~\cite{LiAPL2003}. Raman microscopy
studies have revealed significant phonon confinement effects in
SiNWs down to 4 nm in diameter\cite{AduNanoLett2005}. Very thin
SiNWs with diameters below 5 nm have been synthesized by several
groups \cite{AduNanoLett2005,DDDMa,HolmesScience2000,WuLieberNanoLett2004}
typically oriented in the $\langle100\rangle$,  $\langle110\rangle$,
$\langle111\rangle$, or $\langle112\rangle$  directions.

Most theoretical works on phonon transport in nanowires have been
based on continuous elasticity theory and the Boltzmann transport
equation, e.g. \cite{ZouBalandinJAP2001, SantamorePRL2001}.
Recently, however, some theoretical work based on atomistic models
has been reported: Mingo et al. obtained good agreement with
experiments for wires with diameters $\gtrsim
35\,$nm~\cite{MingoPRB2003,MingoYangNanoLett2003}; Wang et al.
\cite{WangAPL2007} studied the diameter dependence of pristine
$\langle 100\rangle$ wires, and Zhang et al.
\cite{ZhangMingoPRB2007} analyzed the interface scattering between
nanowires and bulk contacts. The effects of an amorphous coating on
the transmission was studied in Ref. \cite{MingoYangPRB2003}.

An important design parameter for SiNW devices is the crystal orientation of the wires, and it is thus highly relevant to compare the thermal transport in the different wire directions. Several recent works have already addressed the directional dependence on the electronic transport properties \cite{Zhao2004,NiquetPRB2005,RuraliPRB2007,VoPRB2006,SvizhenkoPRB2007}. To our knowledge, no experimental comparison of heat conduction in different wire orientations in SiNWs has been published. In wires connected to bulk contacts, Zhang et al.~\cite{ZhangMingoPRB2007} found only insignificant difference between $\langle100\rangle$ and $\langle111\rangle$ wires. Early measurements~\cite{HurstPR1969,McCurdyPRB1970} of anisotropic heat conduction and phonon focusing in bulk silicon was analyzed theoretically~\cite{McCurdy1982} for boundary-scattered phonons in macroscopic samples. However, in very narrow SiNWs with smooth surfaces there will be no surface scattering, since the surface structure is incorporated in the phonon modes. In the ideal ballistic limit, all phonon modes will propagate through the wire with a transmission probability of one.

In this work we compute the thermal conductance of ultra thin
silicon nanowires (SiNWs) with diameters from 1--5 nm oriented along
the $\langle100\rangle$, $\langle110\rangle$, and
$\langle111\rangle$ directions. As a first step we use both DFT
calculations~\cite{siesta-ref} and  empirical
potentials~\cite{Tersoff1988,Gulp} to calculate the dynamical
matrix. Next, the Landauer transmission and thermal conductance are
calculated. We find that the room temperature thermal transport is
highly anisotropic with $\langle110\rangle$ wires having up to two
times larger thermal conductance than $\langle100\rangle$ and
$\langle111\rangle$. Finally, we show that the anisotropies
can be understood based on the underlying  bulk phonon band
structure.

We use atomic orbital DFT calculations~\cite{siesta-ref} to
study the
smallest wires (diameter $\sim 1\,$nm) in the $\langle100\rangle$
and $\langle110\rangle$ directions. After an initial relaxation,
each atom, $I$, is displaced by $Q_{I \mu}$ in direction
$\mu=\{x,y,z\}$ to obtain the forces, $F_{J\nu}(Q_{I\mu})$, on atom
$J\neq I$ in direction $\nu$ \cite{DFTdetails}. The dynamical
matrix, $\mathbf{K}$,  is then found by finite
differences~\cite{FrederiksenPRB2007}
\begin{equation}
K_{I\mu,J\nu}=\frac{\partial^2E}{\partial R_{I\mu}\partial R_{J\nu}}=\frac{F_{J\nu}(Q_{I\mu})-F_{J\nu}(-Q_{I\mu})}{2Q_{I\mu}},
\end{equation}
with $E$ being the total energy. The intra-atomic elements are calculated by imposing momentum conservation, such that $K_{I\mu,I\nu}=-\sum_{K\neq I}K_{I\mu,K\nu}$. The forces are relatively long ranged and we found it necessary to use periodically repeated super cells containing at least 5 unit cells, corresponding to a minimum length $L_{SC}\geq19\,$\AA ~in the $\langle110\rangle$ direction. This implies that the calculations become very time consuming and puts strong limitations on the diameter range to be investigated.

Since the large-diameter DFT calculations become computationally
prohibitive, one needs a faster computational scheme. To this end we
use the Tersoff empirical potential model (TEP)
\cite{Tersoff1988,Tersoff1989} as implemented in the "General
Lattice Utility Program" (\textsc{gulp})\cite{Gulp}. We use
\textsc{gulp} to relax the atomic structure and to directly output
the dynamical matrix, $\mathbf{K}$, for the relaxed system. The
interactions in the TEP are highly local, and the calculations can
be carried out on periodically repeated super cells containing only
two unit cells. The TEP calculations are much faster ($>10^4$ times)
and much less memory demanding than the DFT calculations. In order
to validate the empirical scheme, we present below comparisons
between the DFT and TEP schemes.

Experimentally fabricated SiNWs are typically passivated with either amorphous silicon oxide, SiO$_2$,~\cite{WuLieberNanoLett2004} or with hydrogen~\cite{DDDMa}. The influence of amorphous SiO$_2$ passivation on phonon transport was investigated by Mingo and Yang~\cite{MingoYangPRB2003}. However, atomistic models of amorphous SiO$_2$ are demanding and often one uses hydrogen in calculations of the electronic transport to passivate the silicon dangling bonds~\cite{BlaseNanoLett2006,BlasePRL,MarkussenPRL2007}. The surface passivation is crucial for the electronic properties since the Si dangling bonds form states in the bandgap possibly leading to metallic SiNWs~\cite{RuraliPRL}. Passivation by hydrogen suppresses surface reconstruction and leads to semiconducting wires. We thus include hydrogen when we model SiNWs using DFT calculations. We show below, that it is not necessary to include H-passivation in the TEP model.

\begin{figure}[htb!]
\begin{center}
\includegraphics[width=\columnwidth]{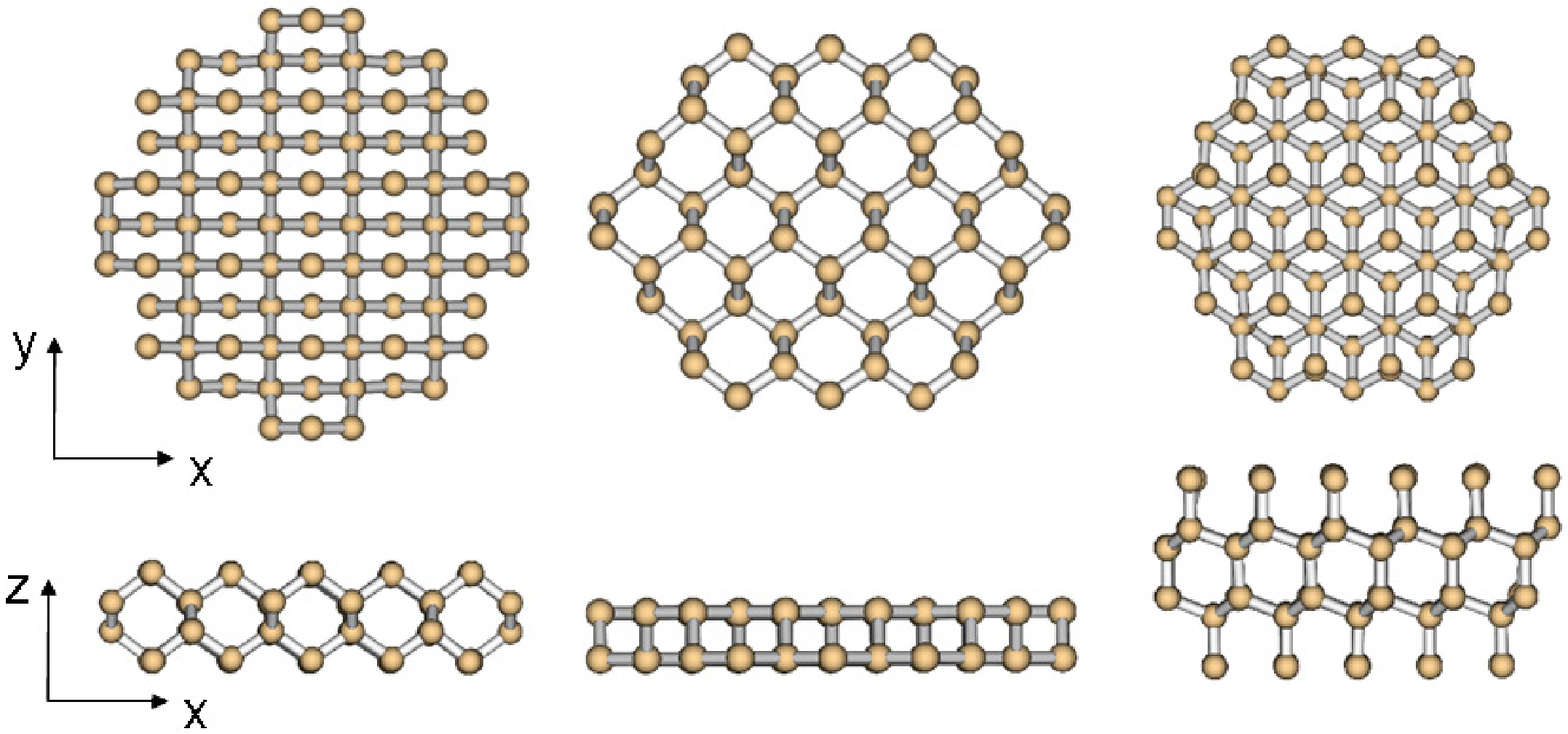}
\includegraphics[width=\columnwidth]{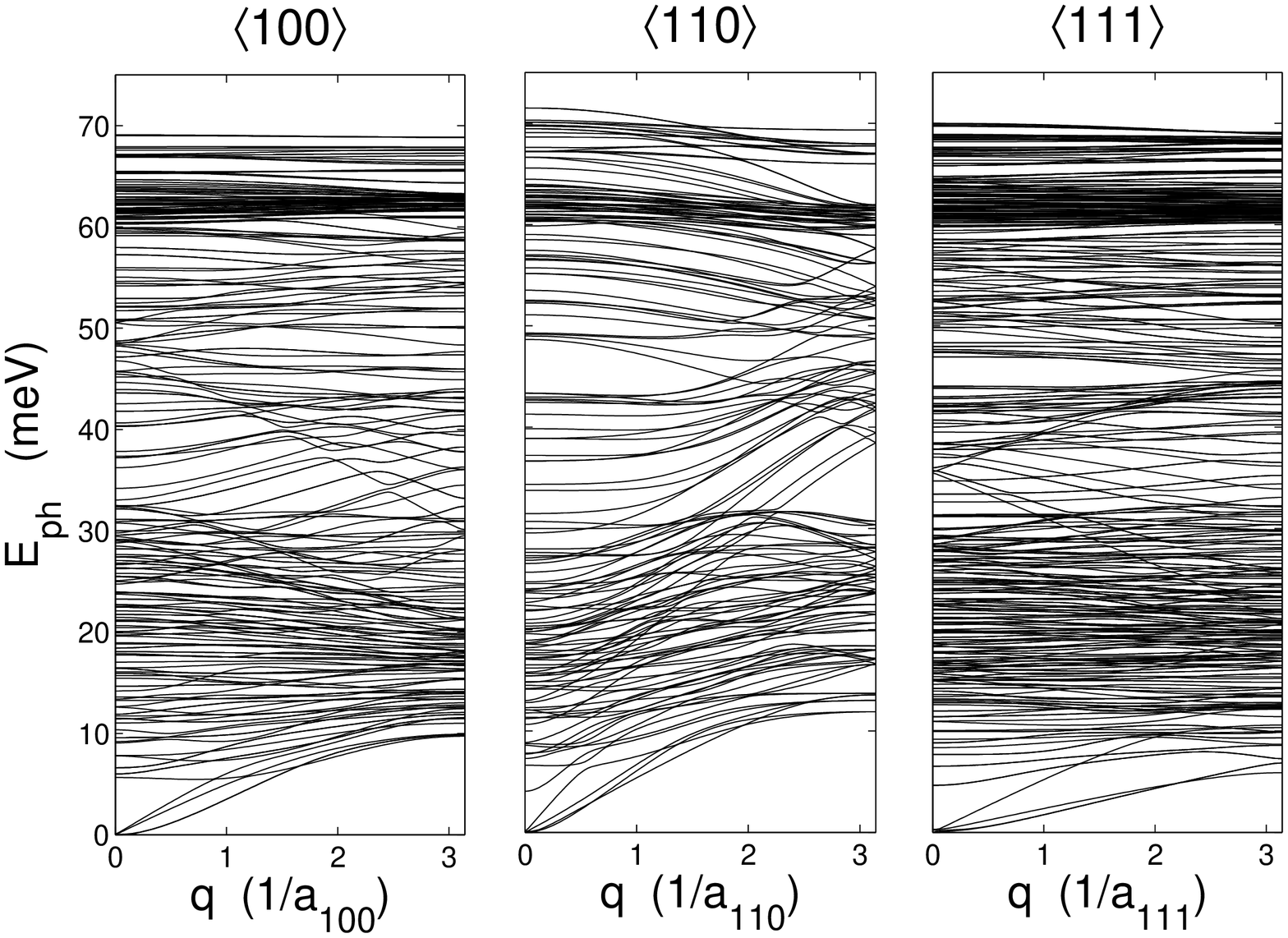}
\end{center}
    \caption{Cross sections (top) and side views (middle) together with the full band structures (bottom) of $\sim$2~nm diameter wires with orientations $\langle100\rangle$ (left), $\langle110 \rangle$ (middle), and $\langle111 \rangle$ (right). The wire is always oriented along the $z-$axis. The phonon wave vectors, $q$ are all in the respective wire directions and are show in units of the reciprocal unit cell lengths, with a$_{100}=5.4\,$\AA, a$_{110}=3.8\,$\AA, and a$_{111}=9.4\,$\AA.    }
\label{wireFig}
\end{figure}

We formally divide our system into a left, central, and right
region. The left and right contacts are modeled as two semi-infinite
pristine wires. In the present study, the central region is also a
pristine wire, but it could in principle include any structural
defects~\cite{MingoYangPRB2003,YamamotoPRL2006,MingoPRB2008} as we
shall describe in a forthcoming paper \cite{vacancyPaper}. In the
limit of a small temperature difference between the left and right
regions, the thermal conductance is
\cite{YamamotoPRL2006,WangPRE2007,MingoPRB2006}
\begin{equation}
G(T) = \frac{\hbar^2}{2\pi k_B T^2}\int_{0}^\infty\mathrm{d}\omega\,\omega^2\,\mathcal{T}(\omega)\,\frac{e^{\hbar\omega/k_BT}}{(e^{\hbar\omega/k_BT}-1)^2}. \label{ThermalConductance}
\end{equation}
$\mathcal{T}(\omega)$ is the transmission function, which in the
ideal ballistic limit equals the number of phonon subbands,
$N_b(\omega)$, at frequency $\omega$ (see Fig.~\ref{wireFig}). In
the low temperature limit we can approximate
$\mathcal{T}(\omega)=\mathcal{T}(0)=4$, since there are four
acoustic modes at $\hbar\omega\rightarrow 0$~\cite{ClelandBook}. The
remaining integral in Eq.(\ref{ThermalConductance}) can be
calculated analytically to yield the universal thermal conductance
quantum $G_Q(T)=4(\pi^2k_B^2 T/3h)$. 

\begin{figure}[htb!]
\begin{center}
\includegraphics[width=.35\textwidth]{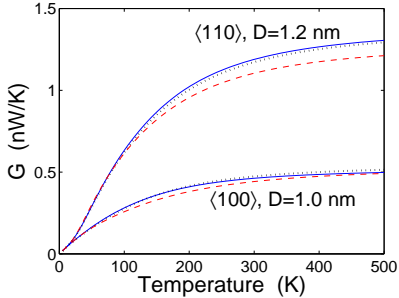}
\end{center}
    \caption{Thermal conductance, vs. temperature for a 1.2 nm diameter $\langle110\rangle$ wire and a 1.0 nm diameter $\langle100\rangle$ wire.  Solid lines: TEP without H-passivation.  Dotted lines: TEP with H-passivation. Dashed lined: DFT.}
\label{gulpSiestaCompare}
\end{figure}

Figure \ref{gulpSiestaCompare} shows the thermal conductance, $G$,
vs. temperature for a 1.2 nm diameter $\langle110\rangle$ wire and a
1.0 nm diameter $\langle100\rangle$ wire.
The results obtained with the TEP model agree quite well with
the microscopic DFT calculations. At temperatures $T<200\,$K, the
relative error is less than 5\%, and at $T=500\,$K the difference is
within 10\%. Also, observe that including hydrogen in the empirical
potential calculations leads to a maximum deviation of 3\% (at
$T=50\,$K, $\langle100\rangle$ wire) as compared to the pure Si
calculations. Leaving out the hydrogen in the TEP calculations is
therefore justified. We take these results as a validation of the
relatively simple and fast empirical potential calculations on pure
Si wires, and proceed to study much larger wires - a study that
would be very time consuming using DFT calculations. In the rest of
this paper, we study the diameter dependence of $\langle100\rangle$,
$\langle110 \rangle$ and $\langle111 \rangle$ oriented SiNWs (see
Figure \ref{wireFig}), using the TEP approach without H-passivation.
We note that the same approximation was recently applied in Refs.
\cite{WangAPL2007,Gutierrez2007}.

Figure \ref{wireFig} shows the atomic structure in the periodically
repeated unit cell (along the $z$-axis) of 2~nm diameter wires with
the three orientations. The top row shows a cross section of the
wires while in the middle row we have shown side views of the unit
cells. In the bottom panel the full phonon band structures for the
three wires are shown. For each wire there are $3N_A$ subbands, with
$N_A$ being the number of atoms in the unit cell. Note that the
different wire orientations have different unit cell lengths,
$a_{hkl}$, and thus also different number of atoms in the unit cell.
At low energies there are only four acoustic modes: the dilatational
and the torsional modes with linear dispersions and two flexural
modes with quadratic dispersions~\cite{ClelandBook}. The many,
relatively flat subbands starting right above the acoustic modes are
denoted shear modes and originate from the bulk acoustic bands that
are folded into the wire axis. The many closely lying bands around
$E_{ph}=63\,$meV correspond to the bulk LO phonon energy.

An important qualitative feature is that the bands in the
$\langle110\rangle$ wire generally have a \textit{larger slope}
(velocity) than the the other wire orientations, where there are
many flat bands. The same trend is also seen for other wire
diameters. Due to the larger slope there are more bands at a given
energy for the $\langle110\rangle$ wires, and one expects
$T_{\langle 110\rangle }>T_{\langle 100 \rangle }\approx T_{\langle
111\rangle }$, leading to a larger thermal conductance for the
$\langle 110\rangle$ wire.

\begin{figure}[htb!]
\includegraphics[width=\columnwidth]{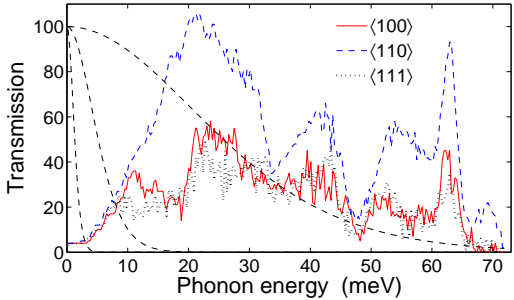}
    \caption{Transmission function for wires with different orientations. The area of the wires are $A_{100}=8.9\,$nm$^2$, $A_{110}=11.2\,$nm$^2$, and $A_{111}=9.0\,$nm$^2$. The three dashed curves show the function $f(\omega,T)$ (see text) at temperatures $T=5,\,20,\,100\,$K (from left to right).}
\label{T_vs_E}
\end{figure}

This anisotropy is further illustrated in Fig.~\ref{T_vs_E} showing
the transmission function for three different wires orientations,
all having diameter $\sim4\,$nm. While the $\langle100\rangle$ and
$\langle111\rangle$ wires have approximately the same transmission,
the $\langle110\rangle$ wire has a significantly larger transmission
at most energies above 10 meV. The three black dashed curves show
the function
$f(\omega,T)=\omega^2\,e^{\hbar\omega/k_BT}/(e^{\hbar\omega/k_BT}-1)^2$
appearing in eq \ref{ThermalConductance} at temperatures
$T=5,\,20,\,100\,$K. We observe that at $T<5\,$K, all wires have
$\mathcal{T}(\omega)=4$, and display the quantized thermal
conductance, $G(T<5K)=G_Q(T)$.

The curve $f(\omega,T=20\,$K) suggests that the thermal conductance
of the three wires should be approximately the same up to
temperatures $T\approx20\,$K. At higher temperatures, the
$\langle110\rangle$ wires will have a higher conductance.

\begin{figure}[htb!]
\includegraphics[width=\columnwidth]{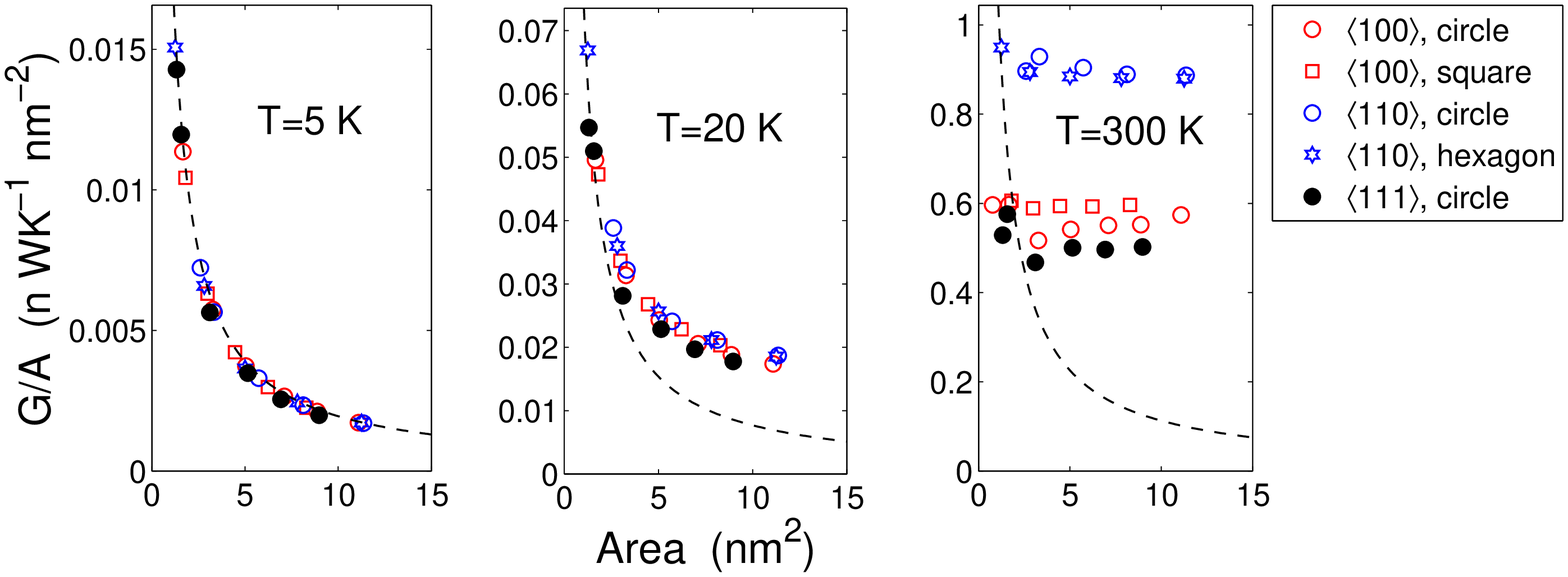}
    \caption{Thermal conductance per unit area, $G/A$, vs. cross sectional area at temperature $T=5\,$K (left), $T=20\,$K (middle), $T=300\,$K (right). The different symbols correspond to different wire orientations as shown in the legend. The dashed lines show $G_Q(T)/A$. Note the different scale in the three plots.}
\label{GvsArea}
\end{figure}

Figure \ref{GvsArea} shows the thermal conductance per unit area, $G/A$, vs.
cross sectional area, $A$, at different temperatures
\cite{howToCalcArea}. The different symbols correspond to different
wire orientations and different cross sectional shapes of the wires.
At low temperatures (5 K, left) all  wires have the same $1/A$
dependence and coincide with the analytical result $G_Q(T)/A$ shown
as the dashed line in the graphs. At $T=20\,$ K (middle), the area
dependence of the three wires is still approximately the same as
expected from Fig.~\ref{T_vs_E}. Note that the thinnest wires still
follow the quantum line, $G_Q(T)/A$. This is because confinement
lifts the bottom of the shear modes (see Fig.~\ref{wireFig}) and
increases the energy range, where only the four acoustic modes
exist.

At $T=300\,$K (right) $G/A$ is almost area
independent. The room temperature area normalized conductances of the
$\langle110\rangle$ oriented wires are approximately 50\% and 75\%
larger than the $\langle100\rangle$ and $\langle111\rangle$ wires,
respectively. The small fluctuations in $G/A$ for a given
orientation are due to variations in the surface structure of the
wires at different diameters. Notice also that for the
$\langle100\rangle$ orientation, the square shaped wires have a
larger conductance than the circular shaped wires. Again, we
attribute this to the details in the surface structure. The room
temperature conductance per unit area of the $\langle100\rangle$ and
$\langle111\rangle$ wires are in fairly good agreement with values
reported in Ref. \cite{WangAPL2007} and \cite{ZhangMingoPRB2007},
respectively.

Experimental~\cite{LiAPL2003} and theoretical~\cite{MingoPRB2003}
works have shown that the thermal conductivity decreases with
decreasing diameter, seemingly contradicting the results shown in
Fig.~\ref{GvsArea}. However, we have not included any scattering
(surface roughnes, impurity, phonon-phonon, phonon-electron) in our
calculations, and our results thus represents an ideal upper limit
for the thermal conductance. Indeed, if we include scattering by
vacancies we observe a decreasing conductance per unit area with decreasing
diameter~\cite{vacancyPaper}. Quantitative comparison with
experiments should only be carried out in the fully ballistic regime
for short wires at relatively low temperatures where anharmonic
effects are unimportant, and for wires with very smooth surfaces.

\begin{figure}[htb!]
\begin{center}
\includegraphics[width=0.8\columnwidth]{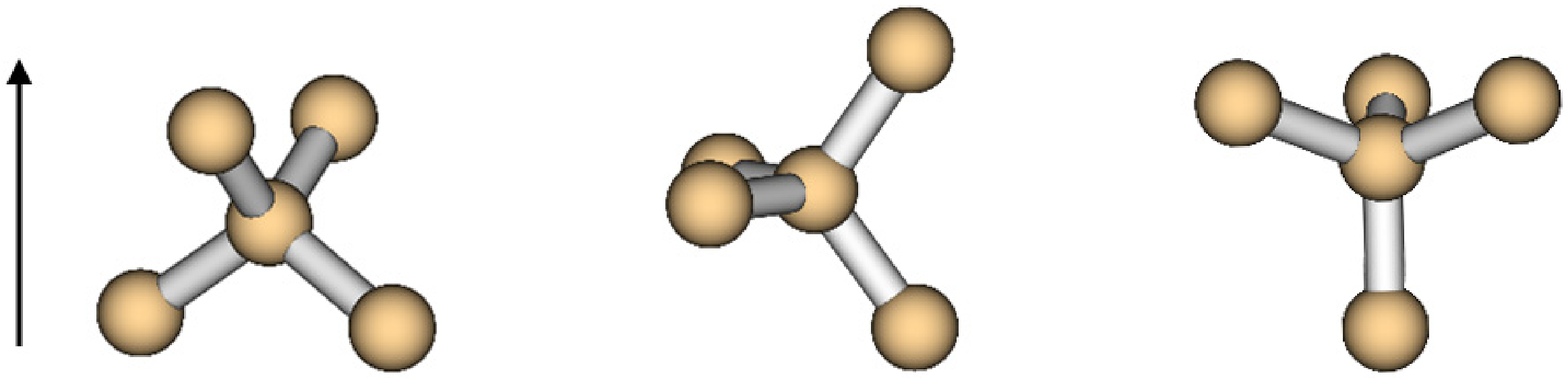}
\includegraphics[width=\columnwidth]{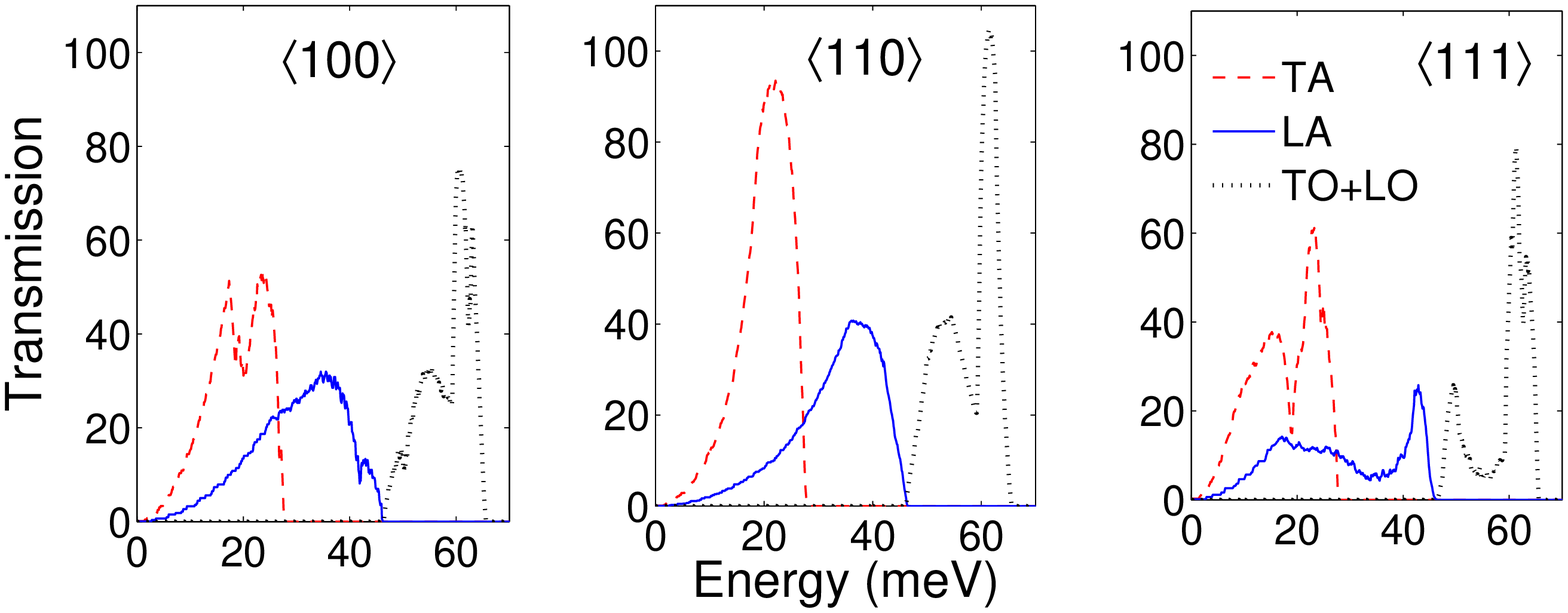}
\end{center}
    \caption{Top panel: Orientation of the basic tetrahedrals for the $\langle100\rangle$, $\langle110\rangle$, and $\langle111\rangle$ directions. The arrow indicate the wire direction as well as the $q_\parallel$ lines. Bottom panel: Transmission functions obtained from the projected bulk band structure. The different curves show the contribution from the two transverse acoustic (TA) modes (dotted red), the single longitudinal acoustic (LA) mode (solid blue) and the three optical (LO+TO) modes (dotted black). The results corresponds to square shaped wires with cross sectional area, $A=9\,$nm$^2$.}
\label{bulkTransmission}
\end{figure}

The anisotropies can also be obtained from the bulk band structure.
We construct an approximate wire band structure by projecting the
bulk subband surfaces onto  lines, $q_\parallel$, along the
$\langle100\rangle$, $\langle110\rangle$, and $\langle 111\rangle$
directions. Figure \ref{bulkTransmission} (top) shows the
orientation of the basic tetrahedrals together with the direction of
the $q_\parallel$ lines. The lines are equidistantly spaced on a
fine square grid in the plane perpendicular to the wire direction,
and are restricted to the first Brillouin zone. A total number of
$N$ lines gives $6N$ projected subbands, since there are two atoms
in the bulk primitive unit cell, and three bands per atom. From the
projected band structure we obtain the transmission by counting the
number of bands at a given energy.  Each of the lines ranges from
$q_\parallel=[0,\pi/a_{hkl}]$, with $a_{hkl}$ being the unit cell
length of a wire in the $\langle hkl\rangle$ direction. For a given
cross sectional area, $A$, the different wire orientations will have
different numbers of atoms in the unit cell and thus also different
number of bands (see Figure \ref{wireFig}). To account for that, we
scale the transmissions by a factor $3N_A/6N$, where
$N_A=\rho\,A\,a_{hkl}$ is the number of atoms in a wire unit cell,
and $\rho=0.05\,$\AA$^{-3}$ is the atomic density.

Figure \ref{bulkTransmission} (bottom panel) shows the transmission functions obtained from the bulk band structure. The large peak in
transmission around $\hbar\omega=20\,$meV for the $\langle 110\rangle$ wire seen in Fig.~\ref{T_vs_E} is reproduced from the bulk bands with the actual transmission values being in close agreement. The room temperature thermal conductances per unit area, $\tilde{G}/A$, obtained from Fig.~\ref{bulkTransmission} are $\tilde{G}_{\langle100\rangle}/A=0.59$, $\tilde{G}_{\langle110\rangle}/A=0.77$, and $\tilde{G}_{\langle111\rangle}/A=0.49\,$nW/(K$\,$nm$^2$), in qualitative agreement with the values in Fig.~ \ref{GvsArea}. Note, however, at low temperature the results are wrong, since the true wires have four acoustic modes, while the projected bulk band structures only have three. At higher temperatures, this error is less important.

Evidently, it is primarily the two TA subbands that are responsible for the large anisotropy. This is not surprising since the bulk band
structure in the $\langle110\rangle$ direction has two non-degenerate TA subbands while in the $\langle100\rangle$ and $\langle111\rangle$ directions the TA subbands are degenerate. Thus, importantly, the anisotropy is not merely a property of the specific silicon nanowires studied here, but can also be expected to take place in other materials with the diamond structure. Indeed, 
projection of the germanium bulk band structure as described above
yields similar results.

In conclusion, we have shown that the ballistic thermal conductance
in silicon nanowires is strongly anisotropic with wires oriented
along the $\langle110\rangle$ direction having 50\% and 75\% larger
conductance per unit area than wires oriented along the $\langle100\rangle$ and
$\langle111\rangle$ directions, respectively. We believe that the
presented results will be relevant for future phonon-engineering of
nanowire devices.

\textbf{Acknowledgement.}
We thank the Danish Center for Scientific Computing (DCSC) and Direkt\o r Henriksens Fond for providing
computer resources. TM acknowledge the Denmark-America foundation for financial support. APJ
is grateful to the FiDiPro program of the Finnish Academy.


\begin{thebibliography}{10}
\bibitem{DFTdetails}
We use a double zeta polarized basis set with an energy cutoff of 100 Ry. The structure is initially relaxed with a maximum residual force of 0.02 eV/\AA. The ions are displaces by $Q_{I \mu}=\pm 0.02\,$\AA.
\bibitem{howToCalcArea}
We define the cross sectional area as $A=N_a/(\rho\, L)$, where $N_a$ is the
  number of atoms in a unit cell, $\rho=0.05\,$\AA$^{-3}$ is the bulk atomic density, and $L$ is the unit cell length.
\end{thebibliography}
\end{document}